\documentclass[12pt, letterpaper, copyedit, double spaced]{article} 

\usepackage{osajnl2}
\usepackage[draft]{hyperref}
\usepackage{amsmath}

\begin{document}

\title{Nonclassicality generated by photon annihilation-then-creation and creation-then-annihilation operations}

\author{Su-Yong Lee,$^{1}$ Jiyong Park,$^1$ Se-Wan Ji,$^{1,2}$ C.H.Raymond Ooi,$^{1,3,4}$ and Hai-Woong Lee$^{1,*}$}

\address{
$^1$Department of Physics, Korea Advanced Institute of Science and
Technology, Daejeon 305-701, Korea
\\
$^2$School of Computational Sciences, Korea Institute for Advanced Study, 207-43 Cheongryangri-dong, Dongdaemun-gu, Seoul 130-012, Korea \\
$^3$School of Engineering, Monash University, Jalan Lagoon Selatan, Bandar Sunway, 46150, Selangor Darul Ehsa, Malaysia\\
$^4$Department of Physics, Korea University, Anam-dong, Seongbuk-gu,
Seoul 136-713, Korea\\
$^*$Corresponding author: hwlee@kaist.edu }

\begin{abstract}We examine nonclassical properties of the field states generated by applying
the photon annihilation-then-creation operation (AC) and
creation-then-annihilation operation (CA) to the thermal and
coherent states. Effects of repeated applications of AC and of CA
are also studied. We also discuss experimental schemes to realize AC
and CA with a cavity system using atom-field
interactions.\end{abstract}


\section{INTRODUCTION}
In many applications of quantum information processing that employ
radiation fields, it is often desirable that the fields possess
nonclassical nature such as entanglement and squeezing. From a
theoretical point of view, perhaps the simplest way of generating
nonclassical field states is to apply the photon creation operation
$\hat{a}^+$ to classical states such as the thermal and coherent
states \cite{Kim}. These nonclassical field states, referred to as
the single-photon added coherent state \cite{Agarwal} and the
single-photon added thermal state \cite{Tara}, respectively, have
been realized experimentally \cite{Zavatta, Zavatta1, Bellini}.

In this paper, we investigate nonclassical properties of the field
states that result when the photon annihilation-then-creation
operation (AC) $\hat{a}^+\hat{a}$ and the photon
creation-then-annihilation operation (CA) $\hat{a}\hat{a}^+$ are
applied to the thermal and coherent states. The noncommutativity of
the photon operations, $[\hat{a},\hat{a}^+]=1$, breaks the symmetry
between AC and CA. It thus is of interest to analyze and compare the
properties of the states that result after AC and CA. In fact, a
direct observation of the noncommutativity of the operators
$\hat{a}$ and $\hat{a}^+$ has been achieved by experimentally
comparing such two states and proving that they differ from each
other \cite{Parigi}. An experimental scheme to directly prove the
commutation relation $[\hat{a},\hat{a}^+]=1$ has also been proposed
recently \cite{Kim1}. Here we look in particular at the possibility
and the degree of nonclassicality exhibited by the states produced
by AC and CA. We also investigate properties of the states obtained
when AC and CA are applied repeatedly to the thermal and coherent
states.

The paper is organized as follows. In Sec. II we briefly review the
effect of the photon creation and annihilation operations
$\hat{a}^+$ and $\hat{a}$. In Sec. III we study properties of the
states produced by AC and CA operated on the thermal state and the
coherent state. We also study the effect of repeated applications of
AC and CA to the thermal and coherent states. Experimental schemes
to realize AC and CA with a cavity system are described in Sec. IV.
Finally, Sec. V presents a discussion of our results.

\section{CREATION AND ANNIHILATION OPERATIONS}
The creation (annihilation) operator $\hat{a}^+$ ($\hat{a}$)
 acting on a photon number state $|n\rangle$ results in an increase (a decrease)
  of the photon number by one. As the creation (annihilation) operator
  is not a Hermitian, however, it cannot be expected to always
  play the role of physically adding (subtracting) one photon. Let us
  consider an arbitrary pure field state $|\psi\rangle=\sum^{\infty}_{n=0}C_n|n\rangle$,
  where $\sum^{\infty}_{n=0}\left|C_n\right|^2=1$.
  The state that results after application of the operator $\hat{a}^+$ ($\hat{a}$)
  to the state $|\psi\rangle$ is $|\psi\rangle_C=N\{\hat{a}^+|\psi\rangle\}
  =\frac{\hat{a}^+|\psi\rangle}{\sqrt{\langle\psi|\hat{a}\hat{a}^+|\psi\rangle}}$
  ($|\psi\rangle_A=N\{\hat{a}|\psi\rangle\}
  =\frac{\hat{a}|\psi\rangle}{\sqrt{\langle\psi|\hat{a}^+\hat{a}|\psi\rangle}}$),
  where $N$ signifies that normalization is to be performed on the unnormalized
  state inside the curly bracket.
  One easily finds
\begin{eqnarray}
\langle\hat{n}\rangle_{C}-\langle\hat{n}\rangle &=& \frac{(\Delta
\hat{n})^2}
{\langle\hat{n}\rangle+1}+1\\
\langle\hat{n}\rangle_{A}-\langle\hat{n}\rangle &=& \frac{(\Delta
\hat{n})^2} {\langle\hat{n}\rangle}-1
\end{eqnarray}
where $\langle\hat{n}\rangle$, $\langle\hat{n}\rangle_C$ and
$\langle\hat{n}\rangle_A$
 refer to the mean photon number of the state $|\psi\rangle$, $|\psi\rangle_C$ and
 $|\psi\rangle_A$, respectively, and $(\Delta \hat{n})^2$ is the photon-number
 variance of the state $|\psi\rangle$ $[(\Delta \hat{n})^2=\langle \hat{n}^2\rangle
 -\langle \hat{n}\rangle^2]$. Eqs. (1) and (2) indicate that, only when the state
 $|\psi\rangle$ has zero photon-number variance, the creation (annihilation) operation
 increases (decreases) the mean photon number by one. The mean photon number
 increases at least by one under the action of the creation operator and decreases at
 most by one under the action of the annihilation operator. In fact, the mean photon
 number can even increase upon application of the annihilation operator,
 which occurs when $(\Delta \hat{n})^2>\langle \hat{n}\rangle$. This
 tendency of the mean photon number to increase beyond one's naive expectation
 has its origin in the relation $\hat{a}^+|n\rangle=\sqrt{n+1}|n+1\rangle$ and
$\hat{a}|n\rangle=\sqrt{n}|n-1\rangle$; a large photon number
carries a greater weight in the state obtained after the $\hat{a}^+$
or $\hat{a}$ operation than in the original state.

The operations $\hat{a}^+$ and $\hat{a}$ can have more profound
effects upon the state they are applied to than simply changing the
number of photons. In fact, the states that result after the
operation $\hat{a}^+$ and $\hat{a}$ often exhibit significantly
different physical properties from the original state and from each
other. It is well known that a creation operation on thermal and
coherent states yields nonclassical states, while the states remain
classical upon an annihilation operation.

\section{ANNIHILATION-THEN-CREATION AND CREATION-THEN-ANNIHILATION OPERATIONS}
In this section we consider the effect of the
annihilation-then-creation operation (AC) $\hat{a}^+\hat{a}$ and the
creation-then-annihilation operation (CA) $\hat{a}\hat{a}^+$ on
thermal and coherent states. We further study the properties of the
states that are obtained when AC and CA, respectively, are applied
repeatedly. We look in particular at the possibility and the degree
of nonclassicality exhibited by these states.

\subsection{Initial thermal state}
The thermal state is characterized by the density matrix
\begin{eqnarray}
\rho=\sum^{\infty}_{n=0}\frac{\overline{n}^n}{(1+\overline{n})^{n+1}}
|n\rangle\langle n|
\end{eqnarray}
where $\overline{n}=\frac{1}{e^{\hbar\omega/kT}-1}$. The mean photon
number $\langle \hat{n}\rangle$ and the photon-number variance
$(\Delta \hat{n})^2=\langle \hat{n}^2\rangle-\langle
\hat{n}\rangle^2$ for the state (3) are given respectively by
$\langle \hat{n}\rangle=\overline{n}$ and $(\Delta
\hat{n})^2=\overline{n}^2+\overline{n}$, yielding the Mandel Q
factor \cite{Mandel} $Q\equiv\frac{(\Delta \hat{n})^2}{\langle
\hat{n}\rangle} -1=\overline{n}$. Application of the operation AC
and CA, respectively, to the state (3) yields
\begin{eqnarray}
\rho_{AC} &=& N\{\hat{a}^+\hat{a}\rho\hat{a}^+\hat{a}\} =
\frac{\hat{a}^+\hat{a}\rho\hat{a}^+\hat{a}}{Tr\{\hat{a}^+\hat{a}\rho\hat{a}^+\hat{a}\}}
\nonumber\\
&=&\frac{1}{(1+2\overline{n})(1+\overline{n})\overline{n}}\sum^{\infty}_{n=0}
\frac{\overline{n}^nn^2}{(1+\overline{n})^n}|n\rangle\langle n| ,\\
\rho_{CA} &=& N\{\hat{a}\hat{a}^+\rho\hat{a}\hat{a}^+\} =
\frac{\hat{a}\hat{a}^+\rho\hat{a}\hat{a}^+}{Tr\{\hat{a}\hat{a}^+\rho\hat{a}\hat{a}^+\}}
\nonumber\\
&=&\frac{1}{(1+2\overline{n})(1+\overline{n})^2}\sum^{\infty}_{n=0}
\frac{\overline{n}^n(n+1)^2}{(1+\overline{n})^n}|n\rangle\langle n|
.
\end{eqnarray}
The Mandel Q factor for the states (4) and (5) can be obtained
through straightforward calculations. We show in Fig. 1 the
calculated Q factor as a function of $\overline{n}$, the mean photon
number of the initial thermal state (3). It can be seen that, when
the AC operation is applied to the thermal state, the Q factor
becomes negative and thus the photon number distribution becomes
sub-Poissonian for a sufficiently small initial mean photon number,
i.e., for $\overline{n} \leq 0.6$. The sub-Poissonian photon
statistics implies nonclassical properties of the state, although
the reverse is not necessarily true. Under the CA operation,
however, the photon statistics remains super-Poissonian regardless
of $\overline{n}$. The figure indicates also that the difference in
the Mandel Q factor between the two states $\rho_{AC}$ and
$\rho_{CA}$ is more significant when $\overline{n}$ is smaller. The
difference in the degree of nonclassicality between the two states
(4) and (5) is illustrated in Fig. 2, in which the corresponding
Wigner distribution functions \cite{Wigner, Lee} are plotted for the
case $\overline{n}=0.57$. Only the Wigner distribution for the state
(4) takes on negative values in the vicinity of the origin.

We now consider the states produced by repeated applications of
$\hat{a}^+\hat{a}$ or $\hat{a}\hat{a}^+$ to the initial thermal
state (3). A straightforward algebra yields
\begin{eqnarray}
\rho^k_{AC} &=& N\{(\hat{a}^+\hat{a})^k\rho(\hat{a}^+\hat{a})^k\}\nonumber\\
&=&\frac{1}{Li_{-2k}(\frac{\overline{n}}{1+\overline{n}})}\sum^{\infty}_{n=0}
\frac{\overline{n}^nn^{2k}}{(1+\overline{n})^n}|n\rangle\langle n|,\\
\rho^k_{CA} &=& N\{(\hat{a}\hat{a}^+)^k\rho(\hat{a}\hat{a}^+)^k\}\nonumber\\
&=&\frac{1}{Li_{-2k}(\frac{\overline{n}}{1+\overline{n}})}\sum^{\infty}_{n=0}
\frac{\overline{n}^{n+1}(n+1)^{2k}}{(1+\overline{n})^{n+1}}|n\rangle\langle
n|,
\end{eqnarray}
where $k$ can be any positive integer and
$Li_s(z)=\sum^{\infty}_{k=1}\frac{z^k}{k^s}$ is the polylogarithm.
The mean photon number ($\langle \hat{n}\rangle^k_{AC}$ and $\langle
\hat{n}\rangle^k_{CA}$) and the variance ($[(\Delta
\hat{n})^2]^k_{AC}$ and $[(\Delta \hat{n})^2]^k_{CA}$) for the
states (6) and (7), can also be obtained by straightforward
calculations. We obtain
\begin{eqnarray}
\langle \hat{n}\rangle^k_{AC} &=&
\frac{Li_{-2k-1}(\frac{\overline{n}}{1+\overline{n}})}
{Li_{-2k}(\frac{\overline{n}}{1+\overline{n}})}=\langle
\hat{n}\rangle^k_{CA}+1,
\end{eqnarray}
\begin{eqnarray}
[(\Delta \hat{n})^2]^k_{AC} &=& [(\Delta \hat{n})^2]^k_{CA}\nonumber\\
&=& \frac{Li_{-2k-2}(\frac{\overline{n}}{1+\overline{n}})}
{Li_{-2k}(\frac{\overline{n}}{1+\overline{n}})}-(\frac{Li_{-2k-1}(\frac{\overline{n}}{1+\overline{n}})}
{Li_{-2k}(\frac{\overline{n}}{1+\overline{n}})})^2.
\end{eqnarray}
We thus see that the mean photon number of state (6) is always
larger by exactly one than that of state (7) (Note, on the other
hand, the operator identity $\hat{a}^+\hat{a}=\hat{a}\hat{a}^+ -1.$)
regardless of $k$. We also see that the photon number variance is
exactly the same for the two states (6) and (7), regardless of $k$.

In general, as the number $k$ of operations $\hat{a}^+\hat{a}$ or
$\hat{a}\hat{a}^+$ is increased from $1$, the mean photon number of
the state increases rapidly. This is due to the fact already
mentioned in Sec. II that a larger photon number carries a greater
weight in the state after the $\hat{a}^+$ or $\hat{a}$ operation
than in the original state. It then is clear from Eqs. (8) and (9)
that, for large $k$, the two states $\rho^k_{AC}$ and $\rho^k_{CA}$
possess almost identical photon-number statistics. In Fig. 3 we plot
the Mandel Q factor as a function of $\overline{n}$ for the two
states $\rho^{k=20}_{AC}$ and $\rho^{k=20}_{CA}$. The two curves are
seen to be very close to each other. Close inspection reveals,
however, that the curve for the state $\rho^{k=20}_{AC}$ lies
slightly below that for the state $\rho^{k=20}_{CA}$, indicating
that the state $\rho^{k=20}_{AC}$ shows slightly stronger
nonclassicality than the state $\rho^{k=20}_{CA}$. For both states,
the Mandel Q factor is negative and the photon-number distribution
is sub-Poissonian if $\overline{n}\leq 0.57$. That the two states
$\rho^{k}_{AC}$ and $\rho^{k}_{CA}$ approach each other as $k$ moves
toward a large number can clearly be seen in Fig. 4, in which the
fidelity between the two states,
$F=\{Tr[\sqrt{\rho^{k}_{AC}}\rho^k_{CA}
\sqrt{\rho^{k}_{AC}}]^{1/2}\}^2$ \cite{Jozsa} is plotted as a
function of $k$, for the case $\overline{n}=0.57$.

\subsection{Initial coherent state}
The coherent state $|\alpha\rangle$ has a Poissonian photon-number
distribution, and thus $Q=0$ for the coherent state. Upon
application of AC and CA, respectively, the coherent state
$|\alpha\rangle$ is transformed to the states
\begin{eqnarray}
|\alpha\rangle_{AC} &=& N\{\hat{a}^+\hat{a}|\alpha\rangle\}=
\frac{\hat{a}^+\hat{a}|\alpha\rangle}{\sqrt{\langle\alpha|
\hat{a}^+\hat{a}\hat{a}^+\hat{a}|\alpha\rangle}},\\
|\alpha\rangle_{CA} &=& N\{\hat{a}\hat{a}^+|\alpha\rangle\}=
\frac{\hat{a}\hat{a}^+|\alpha\rangle}{\sqrt{\langle\alpha|
\hat{a}\hat{a}^+\hat{a}\hat{a}^+|\alpha\rangle}}.
\end{eqnarray}
It should be mentioned that, since the coherent state is an
eigenstate of the annihilation operator $\hat{a}$,
$|\alpha\rangle_{AC}$ is given simply by $|\alpha\rangle_{AC} =
N\{\hat{a}^+|\alpha\rangle\}$ and can be identified as the
single-photon added coherent state.

Fig. 5 shows the Mandel Q factor for the states (10) and (11) as a
function of $\left|\alpha\right|^2$, the mean photon number of the
initial coherent state. It is seen that both states (10) and (11)
have negative Q values regardless of $\left|\alpha\right|^2$. As
before for the case of the initial thermal state, the AC operation
produces stronger nonclassicality than the CA operation, with the
difference more significant when $\left|\alpha\right|^2$ is smaller.
In Fig. 6 we show the Wigner distributions for the states (10)
and(11) for the case $\left|\alpha\right|^2=0.57$. One sees that
both AC and CA operations lead to negative Wigner distributions, but
the degree of negativity created by the AC operation is stronger.

Let us consider the states produced by repeated applications of AC
and CA, respectively, to the coherent state, given by
\begin{eqnarray}
|\alpha\rangle^k_{AC} &=& N\{(\hat{a}^+\hat{a})^k|\alpha\rangle\},\\
|\alpha\rangle^k_{CA} &=& N\{(\hat{a}\hat{a}^+)^k|\alpha\rangle\}.
\end{eqnarray}
Shown in Fig. 7 is the Mandel Q factor for the state
$|\alpha\rangle^{k=20}_{AC}$ and $|\alpha\rangle^{k=20}_{CA}$ as a
function of $\left|\alpha\right|^2$. One sees here that both states
exhibit negative Q factors, but the state
$|\alpha\rangle^{k=20}_{AC}$ shows a slightly stronger
sub-Poissonian photon distribution than
$|\alpha\rangle^{k=20}_{CA}$. Comparing Fig. 5 and Fig. 7, one also
sees that, as $k$ is increased, the two states
$|\alpha\rangle^k_{AC}$ and $|\alpha\rangle^k_{CA}$ approach each
other. In Fig. 8 we plot the fidelity between the two states,
$F=\left|^{k}_{AC}\langle\alpha|\alpha\rangle^k_{CA}\right|^2$, as a
function of $k$, for the case $\left|\alpha\right|^2=0.57$. The
fidelity approaches 1 as $k$ is increased.

\section{PHYSICAL IMPLEMENTATIONS}
The photon creation operation $\hat{a}^+$ on the coherent and
thermal states was realized experimentally by Zavatta et al. using
an optical system based on parametric down-conversion in a nonlinear
crystal followed by conditional detection of a single photon in the idler mode
\cite{Zavatta, Zavatta1, Bellini}. The same group later achieved an experimental realization
of the photon annihilation-then-creation operation (AC)
$\hat{a}^+\hat{a}$ and the photon creation-then-annihilation
operation (CA) $\hat{a}\hat{a}^+$ on the thermal state using again
an optical system, which provided direct observation of the
noncommutativity of the operators $\hat{a}^+$ and $\hat{a}$
\cite{Parigi}. Here we briefly desribe an experimental scheme to
realize AC and CA using a cavity system along the line suggested by
Sun et al. \cite{Sun}.

In order to realize AC, we pass two atoms, atom 1 prepared in the
lower state $|g\rangle$ and atom 2 in the upper state $|e\rangle$,
in order through a cavity. Let us assume an arbitrary pure state
$|\psi\rangle_i=\sum^{\infty}_{n=0}C_n|n\rangle$ for the initial
cavity field. (For simplicity we assume a pure state. Essentially
the same analysis can be applied to obtain the same result even if
the initial cavity field is in a mixed state.) A simple algebra
yields that the state of the cavity field after the interaction of
the cavity with the two atoms, conditioned upon the observation that
atom 1 is found in state $|e\rangle$ and atom 2 in state $|g\rangle$
after the interaction, is given by
\begin{eqnarray}
|\psi\rangle_1=N\{\sum^{\infty}_{n=1}\sin(\sqrt{n}gt_1)\sin(\sqrt{n}gt_2)
C_n|n\rangle\}
\end{eqnarray}
where $g$ is the atom-field coupling constant, and $t_1$ and $t_2$
are, respectively, the interaction time between atom 1 and the
cavity and between atom 2 and the cavity. The state (14) can
approximately be identified as the state
\begin{eqnarray}
|\psi\rangle_{AC}=N\{\hat{a}^+\hat{a}|\psi\rangle_i\}=N\{\sum^{\infty}_{n=1}nC_n|n\rangle\}
\end{eqnarray}
if the interaction times $t_1$ and $t_2$ are sufficiently short that
$\sqrt{\langle \hat{n}\rangle}gt_1\ll 1$ and $\sqrt{\langle
\hat{n}\rangle}gt_2\ll 1$, where $\langle \hat{n}\rangle$ is the
mean photon number of the initial field state $|\psi\rangle_i$.
Repeated operations of AC, $(\hat{a}^+\hat{a})^k$, can then be
realized by letting $k$ pairs of atoms, each pair prepared in
$|g\rangle$ and $|e\rangle$, interact with a cavity, and
postselecting the case where each pair is found in $|e\rangle$ and
$|g\rangle$, subjected to the condition that all interaction times
are sufficiently short.

Similarly, the operation CA can be realized by letting two atoms,
atom 1 prepared in $|e\rangle$ and atom 2 in $|g\rangle$, interact
with a cavity field, conditioned upon atom 1 being found in
$|g\rangle$ and atom 2 in $|e\rangle$ after the interaction and the
interaction times $t_1$ and $t_2$ being sufficiently short. In this
case, the state of the cavity field after the interaction is
\begin{eqnarray}
|\psi\rangle_2=N\{\sum^{\infty}_{n=0}\sin(\sqrt{n+1}gt_1)\sin(\sqrt{n+1}gt_2)
C_n|n\rangle\}
\end{eqnarray}
which, in the short time limit $\sqrt{\langle
\hat{n}\rangle+1}gt_1\ll 1$, $\sqrt{\langle \hat{n}\rangle+1}gt_2\ll
1$, approaches the state
\begin{eqnarray}
|\psi\rangle_{CA}=N\{\hat{a}\hat{a}^+|\psi\rangle_i\}=N\{\sum^{\infty}_{n=0}(n+1)C_n|n\rangle\}.
\end{eqnarray}
Repeated operations of CA, $(\hat{a}\hat{a}^+)^k$, can also be
realized by using $k$ pairs of atoms.

The AC and CA operations can be realized with the cavity scheme
described above with high fidelities $F_1=\left|_1\langle
\psi|\psi\rangle_{AC}\right|^2\cong 1$ and $F_2=\left|_2\langle
\psi|\psi\rangle_{CA}\right|^2\cong 1$, if the interaction times are
sufficiently short. The problem, however,  is that, in the
short-time limit, the success probability of the scheme is low. Let
$P_1$ ($P_2$) be the success probability of the scheme to realize AC
(CA), i.e., let $P_1$ ($P_2$) be the probability to find atom 1 in
$|e\rangle$ ($|g\rangle$) and atom 2 in $|g\rangle$ ($|e\rangle$)
after the interaction. One can easily find that
\begin{eqnarray}
P_1 &=& \sum^{\infty}_{n=1}\sin^2(\sqrt{n}gt_1)\sin^2(\sqrt{n}gt_2)
\left|C_n\right|^2,\\
P_2 &=&
\sum^{\infty}_{n=0}\sin^2(\sqrt{n+1}gt_1)\sin^2(\sqrt{n+1}gt_2)
\left|C_n\right|^2.
\end{eqnarray}
In the short-time limit, $P_1,P_2\propto g^4t^2_1t^2_2 \ll 1$. For
repeated operations $(\hat{a}^+\hat{a})^k$ and
$(\hat{a}\hat{a}^+)^k$, the success probabilities are extremely
small in the short-time limit as $P_1,P_2\propto
g^{4k}t^{2k}_1t^{2k}_2 $.

Instead of taking the short interaction times, one can choose $t_1$
and $t_2$ in such a way that
$\sqrt{\langle\hat{n}\rangle}gt_1=\sqrt{\langle\hat{n}\rangle}gt_2=
\frac{\pi}{2}$ for AC and $\sqrt{\langle\hat{n}\rangle+1}gt_1=
\sqrt{\langle\hat{n}\rangle+1}gt_2= \frac{\pi}{2}$ for CA, where
$\langle \hat{n}\rangle$ is the mean photon number of the initial
cavity field. This choice ensures that the desired state transitions
$|g\rangle \leftrightarrow |e\rangle$ for both atoms 1 and 2 occur
with high probabilities. When the atom makes a transition as
desired, the cavity field gains or loses a photon as desired. One
can thus hope that, with this choice of the interaction times, both
the success probability and the fidelity may stay reasonably close
to unity. Clearly, this strategy would work well if the initial
cavity field has a sharp photon-number distribution centered at
$\langle \hat{n}\rangle$, which is the case for the initial coherent
state but not for the initial thermal state. In Fig. 9 we plot the
success probability $P_1$ and the fidelity $F_1$ we computed as a
function of $\left|\alpha\right|^2$ for the case when the initial
cavity field state is a coherent state, where $t_1$ and $t_2$ are
chosen to satisfy the relation $\sqrt{\langle \hat{n}\rangle}gt_1=
\sqrt{\langle \hat{n}\rangle}gt_2=\frac{\pi}{2}$. One can see
clearly that both the success probability and the fidelity stay
close to unity for $\langle \hat{n}\rangle \geq 50$. The success
probability $P_2$ and the fidelity $F_2$ show similar behavior, the
only significant difference being that, as $\langle \hat{n}\rangle
\rightarrow 0$, $P_1 \rightarrow 0$ as can be seen from Fig 9 but
$P_2 \rightarrow 1$. The result shown in Fig.9 suggests that, with
this choice of the interaction times, even repeated operations of AC
and CA can be realized experimentally with reasonably high success
probability and fidelity, if the initial cavity field has a
sufficiently sharp photon-number distribution.

To  assess  the  feasibility  of  the suggested  experiment,  we  take,  
as  an  example,  $g\cong 10^7 Hz$  and  $\langle \hat{n}\rangle=100$.
 The  interaction  times  $t_1$  and  $t_2$  are  then  required  to  be
  $t_1=t_2=\frac{\pi /2}{\sqrt{\langle \hat{n}\rangle} g}\cong 10^{-8} sec$.
  Since  photon  addition  and  subtraction  should  occur  within  these  
  interaction  times,  we  require  the  spontaneous  emission  decay  rate
  $\gamma$  and  the  cavity  field  decay  rate  $\kappa$  be  small
  compared  with  $\frac{1}{t_1}=\frac{1}{t_2}$, i.e.,  we require
  $\gamma \ll 10^8 Hz $ and $\kappa \ll 10^8 Hz$,  an  experimental
  condition  that  can  be  met  without  too  much difficulty. 
 
 Photon  annihilation-then-creation  and  creation-then-annihilation  operations
 can  be  implemented  experimentally  using  cavity-field  atom  interactions  
 as  described  above  as  well  as  using  optical  means  along  the  line  of
 Refs.\cite{Zavatta, Zavatta1, Bellini}.  The  cavity-QED  method  described  
 here  offers  the flexibility  of
 choosing  interaction  times  in  such  a  way  that  the desired  states
 $|\psi\rangle_{AC}$  and  $|\psi\rangle_{CA}$  can  be  generated  with
 both  high  success  probability  and  high  fidelity.  The  noncommutativity  of
 the  photon  creation  and  annihilation  operations  can  thus  be  verified
 with  high  success  probability  not  easily  achievable  in  the  optical  approach,
 if  the  cavity-QED  method  is adopted.

\section{DISCUSSION}
We have studied effects of the photon annihilation-then-creation
operation (AC) $\hat{a}^+\hat{a}$ and the photon
creation-then-annihilation operation (CA) $\hat{a}\hat{a}^+$, and
their repeated applications on the thermal and coherent states. The
common feature of AC and CA is that they both work to increase the
mean photon number unless the initial state is the number state.
That this is the case can be understood by noting that, due to the
relation $\hat{a}^+\hat{a}|n\rangle=n|n\rangle$ and
$\hat{a}\hat{a}^+|n\rangle=(n+1)|n\rangle$, a larger photon number
carries a greater weight in the state obtained after the operation
$\hat{a}^+\hat{a}$ or $\hat{a}\hat{a}^+$ is applied than in the
initial state. Details of exactly how the weight of each photon
number is changed, however, are different for the two operations.
Close inspection of this difference shed light on one of the main
observations of this work, namely the observation that AC in general
produces stronger nonclassicality than CA. It should be noted, in
particular, that the vacuum state, which is a part of the initial
thermal or coherent state, disappears upon application of the
$\hat{a}^+\hat{a}$ operation, whereas it still remains after the
$\hat{a}\hat{a}^+$ operation. As a result, the state produced by the
$\hat{a}^+\hat{a}$ operation has a significantly different photon
number distribution from the initial ``classical" distribution,
especially if the vacuum state has a large weight in the initial
state, i.e., if the initial thermal or coherent state has a small
mean photon number. This explains why nonclassical properties are
exhibited more strongly after AC than after CA and why the
difference is more pronounced when the mean photon number of the
initial state is small. If the operation $\hat{a}^+\hat{a}$ or
$\hat{a}\hat{a}^+$ is applied repeatedly, the mean photon number of
the state increases and accordingly the weight of the vacuum state
becomes small. In such a case, the effect of the $\hat{a}^+\hat{a}$
operation will not be much different from the effect of the
$\hat{a}\hat{a}^+$ operation. This explains why the two states
produced by applying the $\hat{a}^+\hat{a}$ operation $k$ times and
the $\hat{a}\hat{a}^+$ operation $k$ times approach each other as
$k$ is increased.

In this work, we have used the Mandel Q factor and the Wigner
distribution as an indicator of nonclassicality. In a recent work
\cite{Yang}, nonclassicality of the field states generated by photon
creation-then-annihilation operations were investigated using the
nonclassical depth \cite{Lee2} as a criterion for nonclassicality. We
further note that our investigation here focuses on nonclassical effects
related to sub-Poissonian photon statistics. As photon-added thermal and
coherent states have been shown to exhibit higher-order nonclassical 
effects beyond negativity of the Mandel Q factor and the Wigner
distribution function\cite{Lee3}, the states generated by photon 
creation-then-annihilation operations and by photon annihilation-then-creation
 operations should also exhibit such higher-order nonclassical effects.

\section*{ACKNOWLEDGMENTS} SYL thanks Dr. SunKyung Lee, Dr. Jaeyoon
Cho, and Prof. Suhail Zubairy for helpful discussions. CHRO  acknowledges  support
from  Korea  Research  Foundation  for  the  grant  funded  by  the  Korean 
Government  (KRF-2008-331-C00120). 
 This work was also supported by BK21.


\section*{FIGURE  CAPTIONS}
Fig. 1:
Mandel Q factor of the states $\rho_{AC}$ (dashed curve)
and $\rho_{CA}$ (solid curve) vs. $\overline{n}$, the mean photon
number of the initial thermal state.\\
Fig. 2:
Wigner distribution of (a) the state $\rho_{AC}$ and (b)
the state $\rho_{CA}$ for the case $\overline{n}=0.57$.
$x=\frac{1}{2}(\alpha+\alpha^{\ast})$ and
$y=\frac{1}{2i}(\alpha-\alpha^{\ast})$.\\
Fig. 3:
Mandel Q factor of the states $\rho^{k=20}_{AC}$ (dashed
curve) and $\rho^{k=20}_{CA}$ (solid curve) vs. $\overline{n}$.\\
Fig. 4:
Fidelity between the states $\rho^k_{AC}$ and $\rho^k_{CA}$
vs. $k$ for the case $\overline{n}=0.57$.\\
Fig. 5:
Mandel Q factor of the states $|\alpha\rangle_{AC}$ (dashed
curve) and $|\alpha\rangle_{CA}$ (solid curve) vs.
$\left|\alpha\right|^2$, the mean photon number of the initial
coherent state.\\
Fig. 6:
Wigner distribution of (a) the state $|\alpha\rangle_{AC}$
and (b) the state $|\alpha\rangle_{CA}$ for the case
$\left|\alpha\right|^2=0.57$. $x=\frac{1}{2}(\alpha+\alpha^{\ast})$
and $y=\frac{1}{2i}(\alpha-\alpha^{\ast})$.\\
Fig. 7:
Mandel Q factor of the states $|\alpha\rangle^{k=20}_{AC}$
(dashed curve) and $|\alpha\rangle^{k=20}_{CA}$ (solid curve) vs.
$\left|\alpha\right|^2$.\\
Fig. 8:
Fidelity between the states $|\alpha\rangle^k_{AC}$ and
$|\alpha\rangle^k_{CA}$ vs. $k$ for the case
$\left|\alpha\right|^2=0.57$.\\
Fig. 9:
The success probability $P_1$ (solid curve) and the
fidelity $F_1$ (dashed curve) vs. $\left|\alpha\right|^2$, the mean
photon number of the initial coherent state. The interaction times
$t_1$ and $t_2$ are chosen such that $\left|\alpha\right|
gt_1=\left|\alpha\right| gt_2=\frac{\pi}{2}$.

\begin{figure}[tbp]
\centering
\includegraphics[width=1\columnwidth]{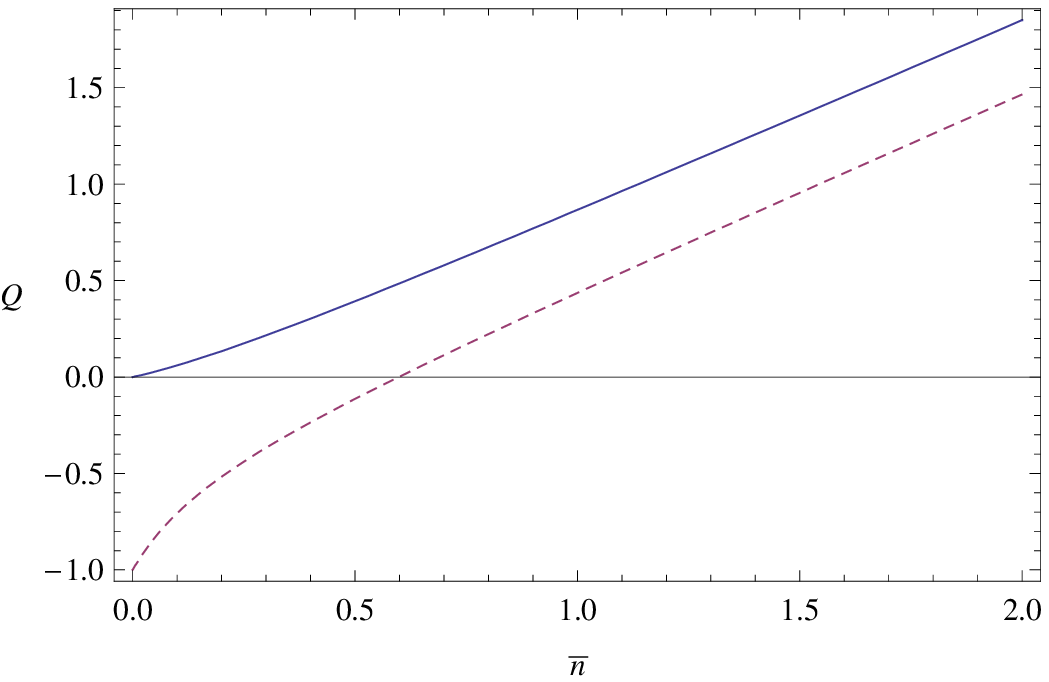}
\caption{}
\end{figure}

\begin{figure}[tbp]
\centering
\includegraphics[width=1\columnwidth]{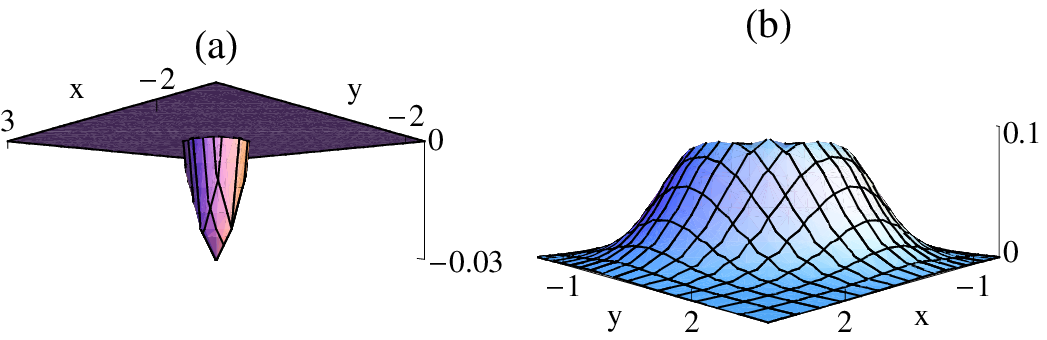}
\caption{}
\end{figure}

\begin{figure}[tbp]
\centering
\includegraphics[width=1\columnwidth]{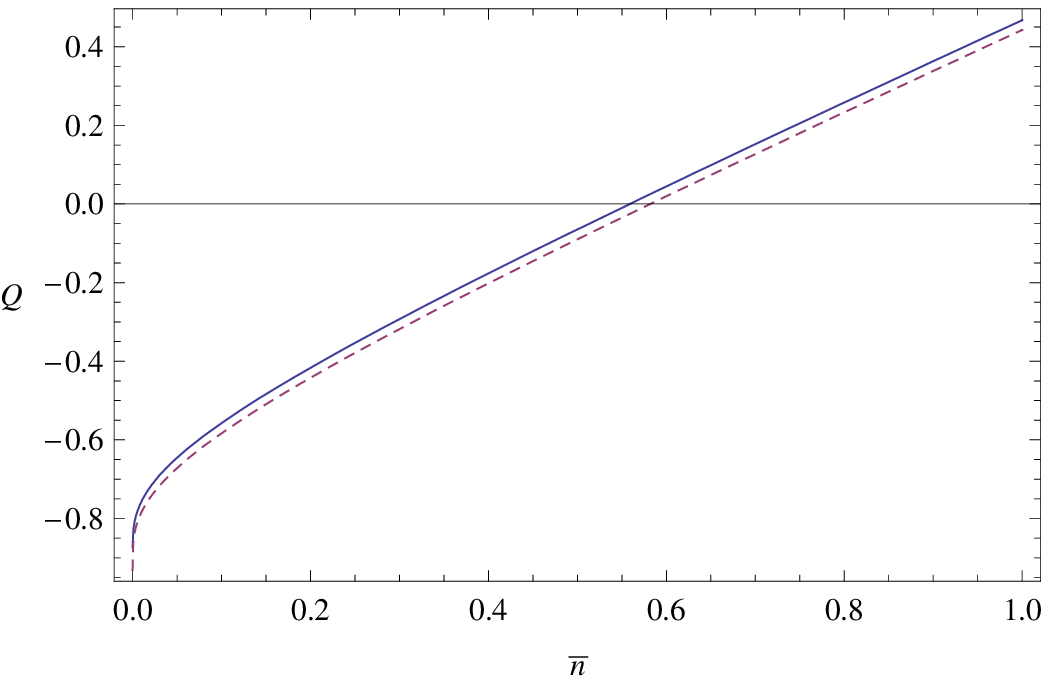}
\caption{}
\end{figure}

\begin{figure}[tbp]
\centering
\includegraphics[width=1\columnwidth]{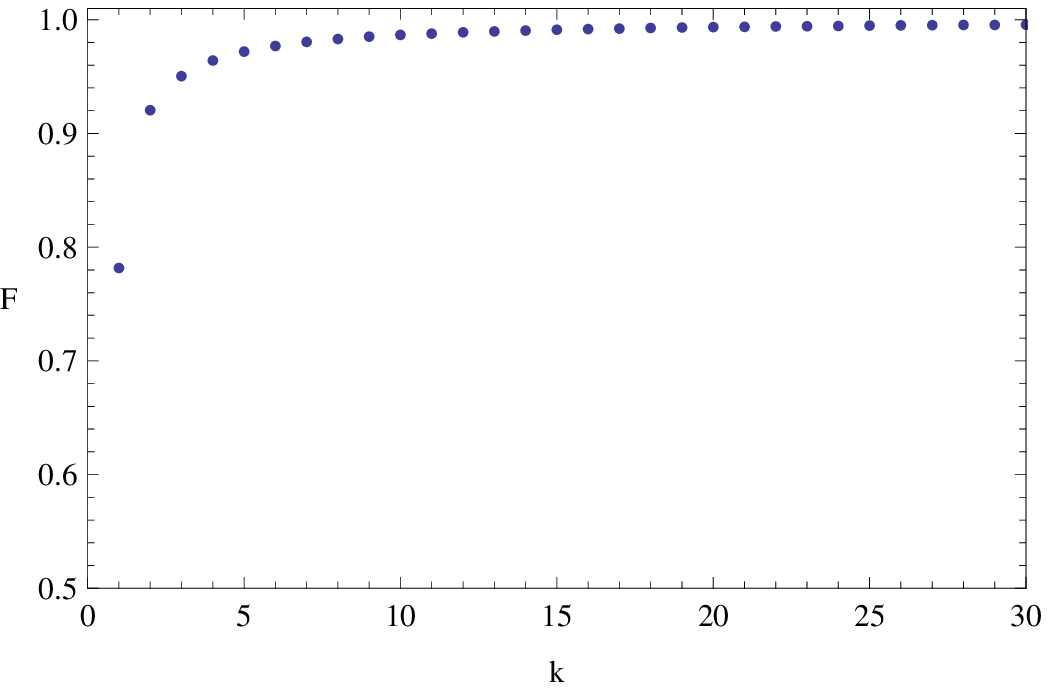}
\caption{}
\end{figure}

\begin{figure}[tbp]
\centering
\includegraphics[width=1\columnwidth]{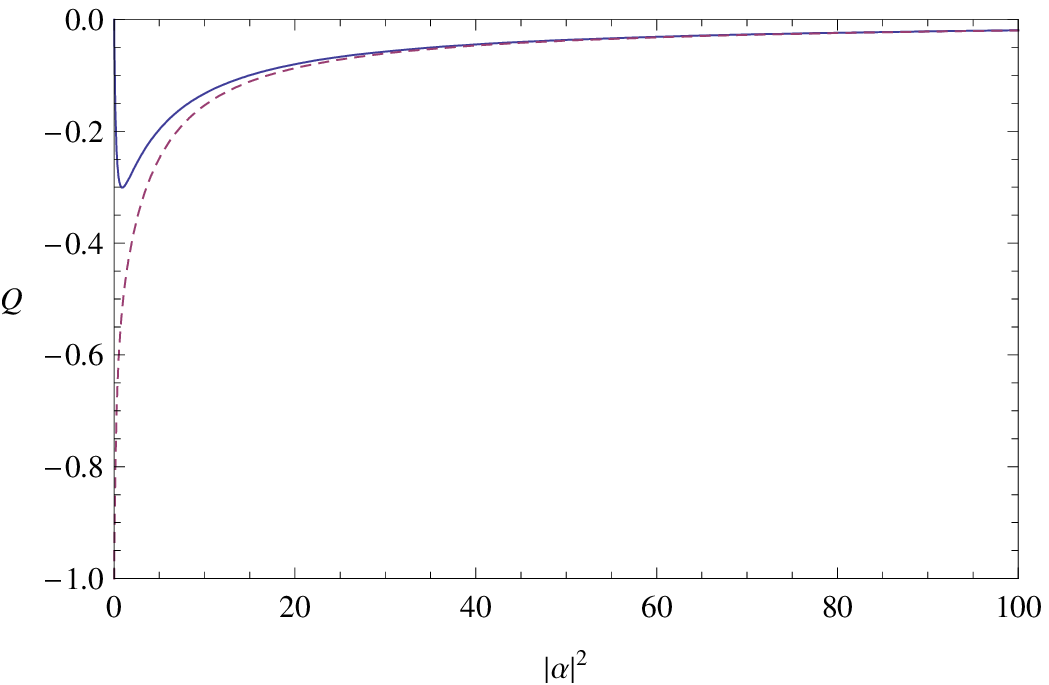}
\caption{}
\end{figure}

\begin{figure}[tbp]
\centering
\includegraphics[width=1\columnwidth]{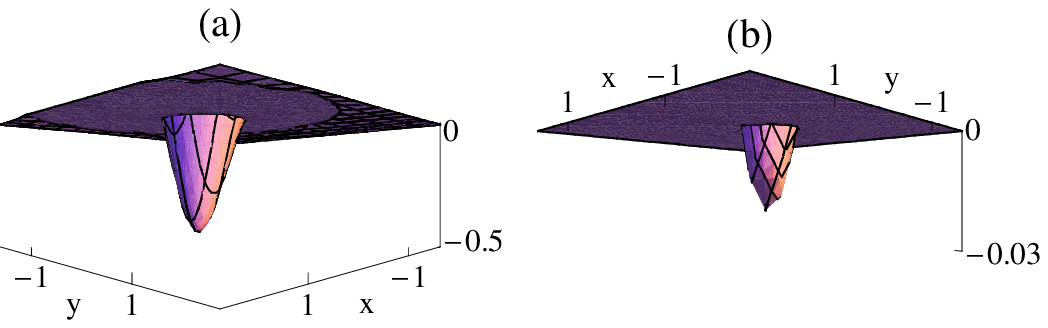}
\caption{}
\end{figure}

\begin{figure}[tbp]
\centering
\includegraphics[width=1\columnwidth]{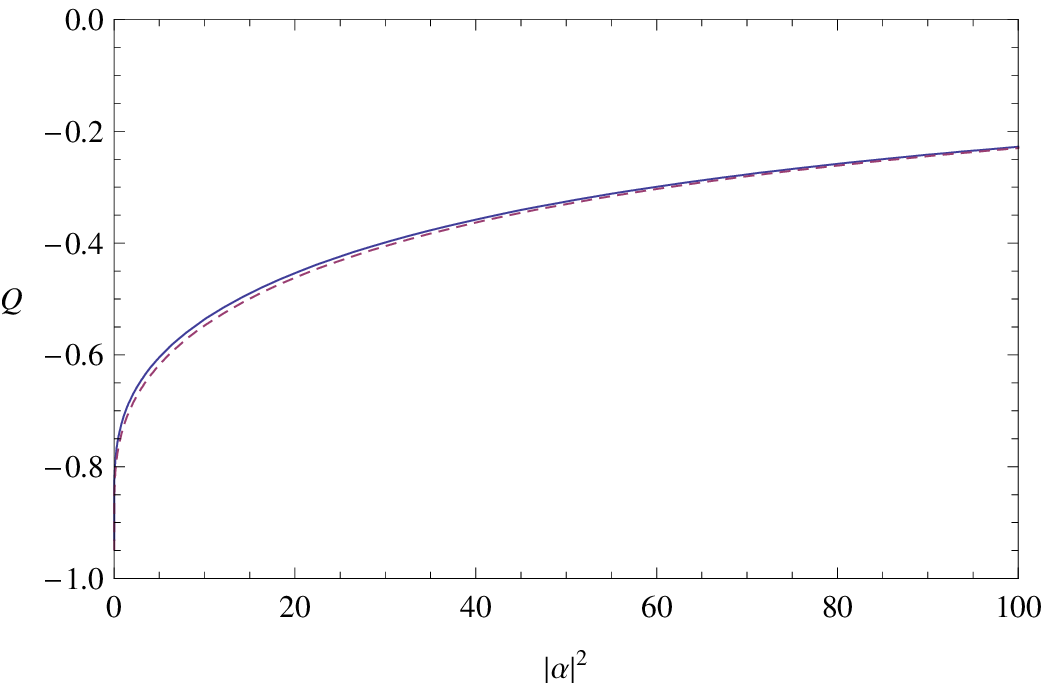}
\caption{}
\end{figure}

\begin{figure}[tbp]
\centering
\includegraphics[width=1\columnwidth]{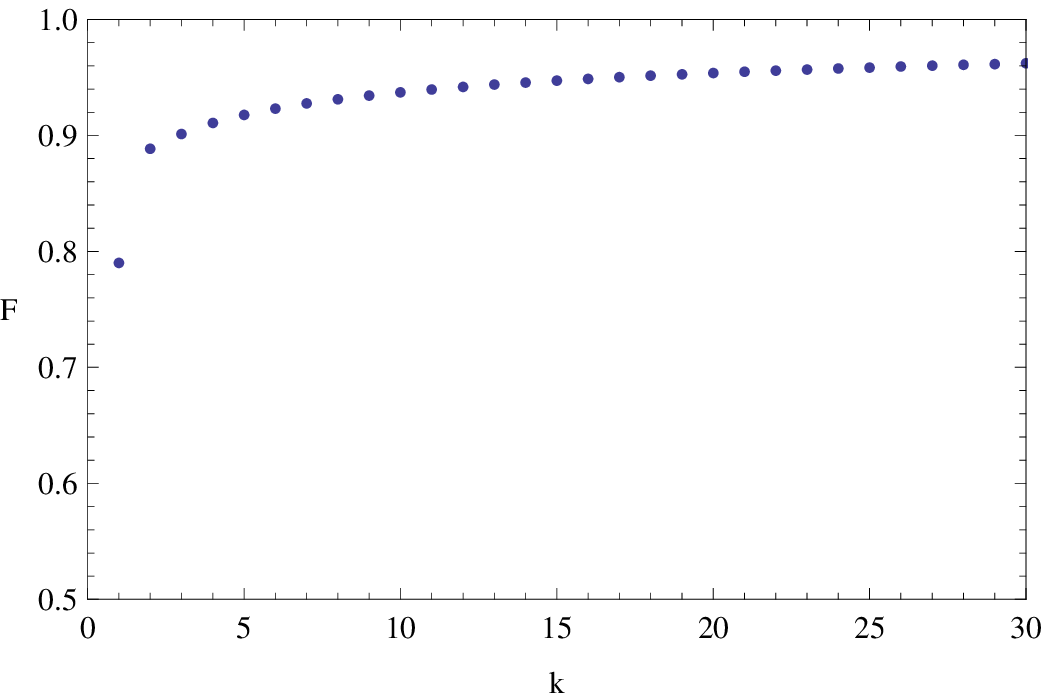}
\caption{}
\end{figure}

\begin{figure}[tbp]
\centering
\includegraphics[width=1\columnwidth]{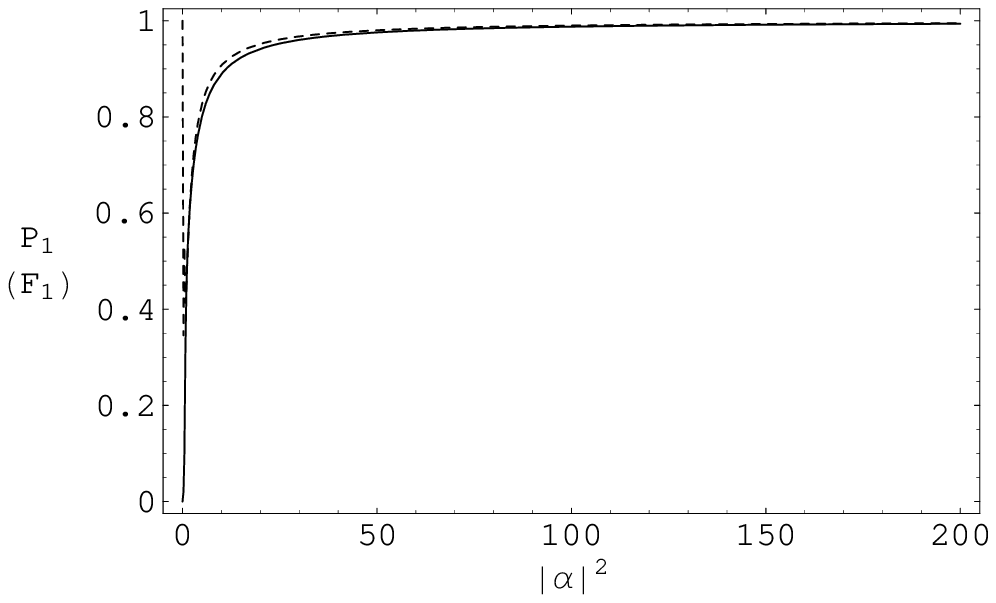}
\caption{}
\end{figure}


\begin{thebibliography}{99}
\bibitem{Kim}  M.S. Kim, ``Recent developments in photon-level operations on travelling light
fields,"  J. Phys. B: At. Mol. Opt. Phys. \textbf{41}, 133001 (2008).

\bibitem{Agarwal}  G.S. Agarwal, and K. Tara,  ``Nonclassical properties of states generated by the excitations on a coherent state,"   \pra {\bf 43}, 492-497 (1991).

\bibitem{Tara}  G.S. Agarwal, and K. Tara, ``Nonclassical character of states exhibiting no squeezing or sub-Poissonian statistics,"  \pra {\bf 46}, 485-488 (1992).

\bibitem{Zavatta}  A. Zavatta, S. Viciani, and M. Bellini, ``Quantum-to-Classical Transition with Single-Photon-Added Coherent States of  Light,"  Science \textbf{ 306}, 660-662 (2004).

\bibitem{Zavatta1}  A. Zavatta, S. Viciani, and M. Bellini, ``Single-photon
excitation of a coherent state: Catching the elementary step of stimulated light emission," 
\pra {\bf 72}, 023820 (2005).

\bibitem{Bellini}  A. Zavatta, V. Parigi, and M. Bellini, ``Experimental nonclassicality of single-photon-added thermal light states,"
 \pra {\bf 75}, 052106 (2007).

\bibitem{Parigi}  V. Parigi, A. Zavatta, M.S. Kim, and M. Bellini,
``Probing Quantum Commutation Rules by Addition and Subtraction of
Single Photons to/from a Light Field," Science \textbf{317},
1890-1893 (2007).

\bibitem{Kim1} M.S. Kim, H. Jeong, A. Zavatta, V. Parigi, and M. Bellini,
``Scheme for Proving the Bosonic Commutation Relation Using
Single-Photon Interference," \prl {\bf 101}, 260401 (2008).

\bibitem{Mandel}  L. Mandel, ``Sub-Poissonian photon statistics in resonance fluorescence,"
 Opt. Lett. \textbf{4}, 205-207 (1979).

\bibitem{Wigner}  E. Wigner, ``On the Quantum Correction For Thermodynamic Equilibrium,"
 Phys. Rev. \textbf{40}, 749-759 (1932).

\bibitem{Lee}  H.W. Lee, ``Theory and application of the quantum phase-space distribution functions," 
Phys. Rep. \textbf{259}, 147-211 (1995).

\bibitem{Jozsa} R. Jozsa, ``Fidelity for mixed quantum states," J. Mod. Opt. \textbf{41}, 2315 -2323(1994).

\bibitem{Sun} Q. Sun, M. Al-Amri, and M.S. Zubairy, ``Probing the quantum commutation rules through cavity QED,"
 \pra {\bf 78},043801 (2008).

\bibitem{Yang} Y. Yang and F.L. Li, ``Nonclassicality of photon-subtracted and photonadded-
then-subtracted Gaussian states," J. Opt. Soc. Am. B \textbf{26},
830-835 (2009).

\bibitem{Lee2} C.T. Lee, ``Measure of the nonclassicality of nonclassical states,"
 Phys. Rev. A, \textbf{44}, R2775-R2778 (1991).
 
\bibitem{Lee3} J.  Lee,  J.  Kim,  and  H.  Nha,  ``Observing  higher-order  nonclassical  effects  using
photon-added  classical  states,"  arxiv:0902.0112[quant-ph].

\end{thebibliography}
\end{document}